%% file: TSR_bias.tex
\documentclass[a4paper,conference,10pt]{IEEEtran}
\usepackage{color}
\usepackage{graphicx}
\usepackage{pstricks}
\usepackage{psfrag}
\usepackage{type1cm}
\usepackage{tabularx}
\usepackage{bbold}
\usepackage[cmex10]{amsmath}
\usepackage{amssymb}
\usepackage{subfig}
\usepackage{tabularx}

\usepackage{algorithmic}
\usepackage[ruled]{algorithm}
\makeatletter
\def\ALG@name{Alg.}

\definecolor{commentgray}{rgb}{.3,.3,.3}

\def\algorithmiccomment#1{\hfill\llap{\hbox to 40.5mm{%
	\color{commentgray}// #1\hss}}}

\makeatother

\def\ve#1{{\mathchoice{\mbox{\boldmath$\displaystyle #1$}}%
              {\mbox{\boldmath$\textstyle #1$}}%
              {\mbox{\boldmath$\scriptstyle #1$}}%
              {\mbox{\boldmath$\scriptscriptstyle #1$}}}}
\def\A{\mathcal{C}}
\def\AO{\mathcal{C}_0}
\def\T{\mathsf{T}}
\def\defeq{\stackrel{\mbox{\scriptsize def}}{=}}
\def\dB{\mathrm{dB}}
\def\SER{\mathrm{SER}}
\def\ul#1{\smash{\underline{\vrule width 0pt height 0pt depth .1pt%
                \smash{\hbox{#1}}}}}
\def\quant{\mathcal{Q}}

\def\argmin{\mathop{\mathrm{argmin}}}

\DeclareSymbolFont{AMSb}{U}{msb}{m}{n}
\DeclareFontFamily{U}{msb}{}%
\DeclareFontShape{U}{msb}{m}{n}{<-6>msbm5<6-8>msbm7<8->msbm10}{}%
\DeclareMathSymbol{\R}{\mathalpha}{AMSb}{"52}
\def\Gauss{\mathcal{N}}

\DeclareSymbolFont{EULE}{U}{eur}{m}{n}
\DeclareMathSymbol{\pdf}{\mathalpha}{EULE}{"66} 
\DeclareMathOperator*{\E}{E}
\DeclareMathOperator*{\var}{Var}
\def\diag{\mathrm{\bf{diag}}}

\def\e{\mathrm{e}}

\def\d{\mathrm{d}}
\def\dx{\,\mathrm{d}x}  
\def\dz{\,\mathrm{d}z}
\def\de{\,\mathrm{d}e}

\def\sigmaN{\sigma_n}

\def\xAPri{\ve{x}_{\mathrm{M}}^{\mathrm{pri}}}
\def\xAPrii{x_{\mathrm{M},i}^{\mathrm{pri}}}
\def\vAPri{\sigma_{{\mathrm{M},\mathrm{pri}}}^2}
\def\xAPost{\ve{x}_{\mathrm{M}}^{\mathrm{post}}}
\def\xAPosti{x_{\mathrm{M},i}^{\mathrm{post}}}
\def\vAPost{\sigma_{\mathrm{M},\mathrm{post}}^2}
\def\vAPosti{\sigma_{\mathrm{M},\mathrm{post},i}^2}
\def\xAExt{\ve{x}_{\mathrm{M}}^{\mathrm{ext}}}
\def\xAExti{x_{\mathrm{M},i}^{\mathrm{ext}}}
\def\vAExt{\sigma_{\mathrm{M},\mathrm{ext}}^2}
\def\vAExti{\sigma_{\mathrm{M},\mathrm{ext},i}^2}
\def\xBPri{\ve{x}_{\mathrm{S}}^{\mathrm{pri}}}
\def\xBPrii{x_{\mathrm{S},i}^{\mathrm{pri}}}
\def\vBPri{\sigma_{\mathrm{S},\mathrm{pri}}^2}
\def\xBPost{\ve{x}_{\mathrm{S}}^{\mathrm{post}}}
\def\xBPosti{x_{\mathrm{S},i}^{\mathrm{post}}}
\def\vBPost{\sigma_{\mathrm{S},\mathrm{post}}^2}
\def\vBPosti{\sigma_{\mathrm{S},\mathrm{post},i}^2}
\def\xBExt{\ve{x}_{\mathrm{S}}^{\mathrm{ext}}}
\def\xBExti{x_{\mathrm{S},i}^{\mathrm{ext}}}
\def\vBExt{\sigma_{\mathrm{S},\mathrm{ext}}^2}

\def\c{\bar{c}}
\def\PhiDD{\ve{\Phi}_{dd}}

\def\xUi{x_{\mathrm{U},i}}
\def\sigmaUi{\sigma_{\mathrm{U},i}^2}
\def\W{\mathcal{W}}

\IEEEoverridecommandlockouts 

\begin{document}
\title{\vspace*{-2mm} Unveiling Bias Compensation in Turbo-Based Algorithms for (Discrete) Compressed Sensing}
\author{\thanks{This work was supported by Deutsche
                Forschungsgemeinschaft (DFG) under grant FI~982/8-1.}%
		Susanne Sparrer, Robert F.H. Fischer \\
		Institute of Communications Engineering,
		Ulm University, 89081 Ulm, Germany \\
		Email: susanne.sparrer@uni-ulm.de, robert.fischer@uni-ulm.de \vspace*{-1.5mm}}
\maketitle

\begin{abstract}
In Compressed Sensing, a real-valued sparse vector has to be recovered 
from an underdetermined system of linear equations. In many applications, 
however, the elements of the sparse vector are drawn from a finite set. Adapted 
algorithms incorporating this additional knowledge are required for the 
discrete-valued setup. In this paper, turbo-based algorithms for both cases 
are elucidated and analyzed from a communications engineering perspective, 
leading to a deeper understanding of the algorithm. In particular, 
we gain the intriguing insight that the calculation of extrinsic values 
is equal to the unbiasing of a biased estimate and present an improved 
algorithm.
\end{abstract}

\input{TSR_bias_c1}

\input{TSR_bias_c2}

\input{TSR_bias_c3}

\input{TSR_bias_c4}

\input{TSR_bias_c5}


\input{TSR_bias_a1}

\end{document}

%% file: TSR_bias_c1.tex
\section{Introduction}
\label{sec_1}

\noindent
In many communication scenarios, the transmitted vector 
$\ve{x}\in\R^{L\times 1}$ is \emph{sparse}, i.e., only 
$s$ elements are non-zero.%
\footnote{Notation: $||\cdot||_p$ denotes the $\ell_p$ norm. 
	$\ve{A}_{(l,m)} = A_{l,m}$ is the element in the $l^\mathrm{th}$ row and
	$m^\mathrm{th}$ column of $\ve{A}$. 
	$\ve{A}^{\T}$ and $\ve{A}^{-1}$ denote the transpose and the inverse 
	of $\ve{A}$, respectively.
	$\diag(\ve{a})$ denotes a diagonal matrix of appropriate size with
	entries of the vector $\ve{a}$ as  diagonal elements.
	$\diag(\ve{A})$ denotes a diagonal matrix with the same diagonal 
	elements as $\ve{A}$.
	$\ve{I}$ is the identity matrix.
	$\quant_{\A}(\cdot)$: element-wise quantization w.r.t.\ a given alphabet
	$\A$.
	$\E\{\cdot\}$: element-wise expectation.
	$\var\{\cdot\}$: Variance.
	$\Gauss(m,v)$: Gaussian distribution with mean $m$ and variance $v$.
	$\pdf_X(x)$: probability density function of random variable $x$.
} 
The receive vector 
$\ve{y}\in\R^{K\times 1}$, $s \ll K < L$, is corrupted by i.i.d.\ zero-mean 
Gaussian noise $\ve{n}$, with variance $\sigmaN^2$ per component.
The channel can then be modeled by
\vspace*{-1mm}	
\begin{align}
  \ve{y} = \ve{A} \ve{x} + \ve{n} \ , \label{eq_channel_model}
\end{align}
\vspace*{-5mm}	

\noindent
where $\ve{A}\in\R^{K \times L}$ corresponds to the channel matrix 
in terms of communications engineering.
Since $K< L$, a sparse vector has to be recovered from an 
underdetermined system of linear equations, a problem which is known as
\emph{\ul{C}ompressed \ul{S}ensing} (CS) \cite{Donoho:2006}. In many 
communication scenarios, however, the non-zero elements are not real-valued 
but drawn from a finite set $\A$. 
If the sparsity is fixed, the problem to be solved is given by 
($\AO \defeq \A \cup \{0\}$)
\vspace*{-2mm}
\begin{align}
  \hat{\ve{x}} = \argmin_{\tilde{\ve{x}} \in\AO^L} \|\ve{y}-\ve{A}\tilde{\ve{x}}\|_{2}^2
	\quad \text{s.t.} \quad \|\tilde{\ve{x}}\|_{0} = s \ . 		\label{eq_problem_sKnown}
\end{align}
\vspace*{-3mm}	

\noindent
There are many fields of digital communication in which this problem 
of estimating a \emph{discrete-valued sparse} vector from an 
\emph{underdetermined} system of linear equations is present, 
such as, e.g., sensor networks, where a fusion center with $K$ 
antennas has to reconstruct which of the $L$ low-activity sensors 
have currently been active, and which data has been transmitted 
by them \cite{Zhu:11}. 
Further applications are peak-to-average 
power reduction in orthogonal frequency-division multiplexing 
\cite{Fischer:12}, the detection of pulse-width-modulated 
signals in radar applications \cite{Ens:13},  
\ul{c}ode-book \ul{e}xcited \ul{l}inear \ul{p}rediction 
(CELP) source coding \cite{Dymarski:13}, and 
Compressed-Sensing-based cryptography \cite{Fay:16}.

There is a tremendously wide range of algorithms solving the 
standard continuous-valued CS problem, such as, amongst others, 
\ul{O}rthogonal \ul{M}atching \ul{P}ursuit (OMP) \cite{Pati:93}, 
\ul{I}terative \ul{H}ard \ul{T}hresholding (IHT) \cite{Blumensath:08}, and 
\ul{I}terative \ul{S}oft \ul{T}hresholding (IST) \cite{Daubechies:08}. 
Although the standard CS problem is non-convex due to its sparsity 
constraint, it can be relaxed to a $\ell_1$-based problem, which can 
efficiently be solved by the simplex algorithm or interior point methods 
\cite{Nemhauser:88}.

In the case of \emph{discrete Compressed Sensing}, however, 
additional information, i.e., the knowledge that the elements of 
the sparse vector are from a finite set, is available and has 
to be taken into account adequately.
The estimation of a discrete-valued vector has combinatorial complexity 
in general, and hence discrete CS is non-convex, even if the constraint 
in problem (\ref{eq_problem_sKnown}) was relaxed to an $\ell_1$-based one, 
which could be solved by extensions of the simplex algorithm 
\cite{Nemhauser:88}. Unfortunately, these algorithms have a prohibitively 
high computational complexity.

Some algorithms for the solution of problem (\ref{eq_problem_sKnown}) have 
been proposed over the last few years. Besides the most obvious approach 
of a standard CS algorithm with subsequent quantizer \cite{Sparrer:14}, 
the quantization can be included inside OMP \cite{Sparrer:15}, 
which equals the so-called model-based Compressed 
Sensing \cite{Baraniuk:10} if it is applied to discrete CS. This algorithm 
has been further improved by the application of a method which preserves 
reliability information \cite{Sparrer:15}. 
Another improved variant of OMP has been introduced in \cite{Sparrer:15b}, 
where a minimum mean-squared error estimator has been applied.

Other algorithms for the CS problem are related to well-known channel decoding 
algorithms, e.g., the \ul{a}pproximate \ul{m}essage-\ul{p}assing (AMP) algorithm 
\cite{Donoho:10, Bayati:11}, which is derived from the message-passing algorithm 
\cite{Kschischang:01} and which can be easily adapted to cases (such as discrete CS) 
where information on the a-priori distribution of the sparse vector is available (\ul{B}ayesian AMP, BAMP) 
\cite{Donoho:10}. Moreover, the knowledge from channel coding has also been used for 
the optimization of measurement matrices with adapted specialized recovery algorithms, 
cf., e.g., \cite{Dai:08a, Dai:08b}. The drawback of these approaches is that the 
restrictions on the measurement matrix limit the range of applications.

In \cite{Ma:15a,Ma:15b}, an approach which is based on the turbo-principle has been 
proposed. It has been simplified, generalized, and adapted to the discrete setup 
in \cite{Sparrer:17}. 
In this paper, this algorithm is further improved, thereby including knowledge 
from the field of digital communications into CS. The aim of this paper is to 
gain a profound understanding of the algorithm. An analysis of the approaches 
leads to intriguing insights into the algorithm, and especially into the 
important topic of bias compensation. 
Note that the results hold for standard CS and for discrete CS.

The paper is organized as follows. In Sec.~\ref{sec_2}, the improved algorithm is 
introduced. The analysis is given in Sec.~\ref{sec_3}. In Sec.~\ref{sec_4}, the 
performance of the algorithms is compared, followed by brief conclusions in 
Sec.~\ref{sec_5}.

%% file: TSR_bias_c2.tex
\def\mycolorbox#1#2{\mbox{}\hskip-\fboxsep\colorbox{#1}{#2}}

\section{Turbo Signal Recovery}
\label{sec_2}

\noindent
In \cite{Ma:15a,Ma:15b}, an iterative algorithm for the estimation of 
complex-valued sparse vectors has been presented. 
This algorithm, which has been denoted as \ul{T}urbo \ul{S}ignal \ul{R}ecovery 
(TSR) (or Turbo Compressed Sensing), has been restricted to one 
type of measurement matrices in the original work. In \cite{Sparrer:17}, the 
algorithm has been generalized to a wider range of matrix construction, which allowed 
the comparison of TSR with other algorithms. Furthermore, the notation of the 
algorithm has been simplified. In this section, after a short explanation of this 
modified algorithm, an improved version is introduced. 

\subsection{Approximate LMMSE TSR}
\noindent
In the generalized TSR algorithm, the measurement matrix is assumed to be given by 
$\ve{A}=\ve{U}\ve{C}$, where $\ve{U}$ is a random part of a unitary matrix, and 
$\ve{C} = \diag([c_1,\ldots,c_L])$ is a scaling matrix. 
In the standard CS setup where the \emph{column} vectors of $\ve{A}$ are assumed to 
be normalized to unit length, the scaling factors calculate to 
$c_i^2 = 1/\sum_{j=1}^{K} U_{j,i}^2$. 

The sparse vector is estimated in an iterative fashion, where each iteration consists 
of two main parts: In the first part, all elements of the vector are estimated jointly by 
a linear estimator aiming to keep the Euclidean estimation error small, thereby ignoring 
the sparsity and the alphabet constraint. In the second part, the constraints which have 
been disregarded in the first part are taken into account, which leads to a non-linear 
estimator which generates so-called soft values \cite{Tarkoey:95}. The pseudocode 
of this algorithm is given in Alg.~\ref{Algo_TurboCS_withoutZ}, Variant A, i.e., only the 
lines tagged by an ``A'' are active. 

The authors of the original TSR algorithm denote the first step as \ul{l}inear 
\ul{m}inimum \ul{m}ean-\ul{s}quare \ul{e}rror (LMMSE) estimation \cite{Ma:15a}. 
In this step, the sparse vector is estimated by (cf.\ Alg.~\ref{Algo_TurboCS_withoutZ}, 
Line \ref{algATSR_LineMMSE})
\vspace*{-1mm}
\begin{align}
  \xAPost = \xAPri + \frac{\c^2\vAPri}{\c^2 \vAPri + \sigmaN^2} \cdot \ve{C}^{-1}\ve{U}^{\T} (\ve{y} - \ve{A}\xAPri) \ , \label{eq_ATSR_firstStep}
\end{align}
\vspace*{-3mm}	

\noindent
where $\xAPri$ is a prior estimate (from the previous step), $\vAPri$ is the variance 
of the estimation error of this prior estimate, and $\c^2 = \frac{1}{L} \sum_{i=1}^L c_i^2$ 
is the average scaling factor. All variables of this \ul{M}MSE-estimation-step are 
marked by the index ``$\mathrm{M}$''.

In \cite{Sparrer:17}, it has been shown that with $\ve{C} \approx \c\ve{I}$, 
(\ref{eq_ATSR_firstStep}) can be approximated by
\vspace*{-1mm}
\begin{align}
  \xAPost \approx \xAPri + \frac{\vAPri}{\c^2 \vAPri + \sigmaN^2} \cdot \ve{A}^{\T} (\ve{y} - \ve{A}\xAPri) \ . \label{eq_ATSR_firstStepApprox}
\end{align}
\vspace*{-3mm}	

\noindent
Note that (when ignoring the scaling factor) this term corresponds to the first 
step in the well-known IHT algorithm \cite{Blumensath:08}, which can, on the one 
hand, be interpreted as one step of the gradient descent method, but, on the other 
hand, as correlation-based estimation, since $\ve{x}$ is estimated based on the 
correlation between the residual and the column vectors of $\ve{A}$. From a 
communications engineering point of view, it corresponds to the application of a 
matched filter. Hence, although claimed otherwise by the authors in 
\cite{Ma:15a}, the first step is not an LMMSE estimation.

In the second step, soft values are calculated, which are the expected 
value of $x$ conditioned to the prior estimate from the first step and the a-priori 
distribution of $x$ (cf.\ Alg.~\ref{Algo_TurboCS_withoutZ}, Line \ref{algTSR_LineSF}).
Note that, in contrast to the joint estimation in the first step, this calculation 
is performed for each element of the sparse vector individually. Since this approach 
takes the a-priori distribution of $x$ into account, it depends on the alphabet;  
an adaptation to any alphabet is straightforward. This approach is also used in other 
algorithms for (discrete) CS, cf., e.g., \cite{Donoho:10, Sparrer:17, Sparrer:16, Sparrer:15}.
All variables of the second (\ul{s}oft-value calculating) step are indicated by the 
index ``$\mathrm{S}$''.

This algorithm is denominated as TSR/Q in the following, 
where the trailing ``Q'' emphasizes the final quantization step which has 
to be performed in order to restrict the estimate to the discrete alphabet.

\def\algorithmiccomment#1{\rlap{\hspace*{0.7cm}\color{commentgray}// #1 \hss}}
\makeatletter
\def\AlgLine#1#2{%
	\def\ALC@lno{\ALC@linenosize \arabic{ALC@line}%
		\ifnum#1=1 A\else\phantom{A}\fi%
		\ifnum#2=1 B\else\phantom{B}\fi%
		\ALC@linenodelimiter}%
	}
\def\AlgDC{\addtocounter{ALC@line}{-1}}
\def\AlgSC#1{\setcounter{ALC@line}{#1}}
\def\mycolorbox#1#2{\mbox{}\hskip-\fboxsep\colorbox{#1}{#2}}
\makeatother
\begin{algorithm}
\caption{\label{Algo_TurboCS_withoutZ}
  $\hat{\ve{x}} = \mathrm{recover} \left(\ve{y}, \ve{U}, \ve{C}, \sigmaN^2, s, \AO \right)$ \newline
  Variants: A: TSR/Q, B: TMS/Q
}
\small{
\algsetup{indent=.5em}
\begin{algorithmic}[1]
\AlgLine{1}{1}
\STATE $\xAPri = \ve{0}$, $\vAPri = s/L$, $\ve{A} = \ve{U} \ve{C}$
\AlgLine{1}{1}
\WHILE{stopping criterion not met}
\AlgLine{0}{0} 

\COMMENT{MMSE estimation}
\AlgLine{1}{0}
  \STATE $\xAPost = \xAPri + \frac{\c^2\vAPri}{\c^2 \vAPri + \sigmaN^2} \cdot \ve{C}^{-1}\ve{U}^{\T} (\ve{y} - \ve{A}\xAPri) $
      \label{algATSR_LineMMSE}
\AlgDC
\AlgLine{0}{1}
  \STATE $\xAPost  \hspace*{-.25em} =  \hspace*{-.2em} \xAPri \hspace*{-.12em} + \vAPri \ve{A}^{\T} \left(\vAPri \ve{A}\ve{A}^{\T}  \hspace*{-.12em} + \hspace*{-.12em} \sigmaN^2 \ve{I}\right)^{-1} \hspace*{-.22em} (\ve{y} - \ve{A}\xAPri) $ 
      \label{algETSR_LineMMSE}
\AlgLine{1}{0}
  \STATE $\vAPost = \vAPri \cdot \left(1 - \frac{K}{L} \, \frac{\c^2\vAPri}{\c^2\vAPri + \sigmaN^2} \right)$
\AlgDC
\AlgLine{0}{1}
\STATE $\ve{K} = \vAPri \ve{A}^{\T} \left(\vAPri \ve{A}\ve{A}^{\T} + \sigmaN^2\ve{I}\right)^{-1} \hspace*{-.5em}\ve{A}$
\AlgLine{0}{1}
  \STATE $\vAPost = \vAPri \cdot \left( 1 - \frac{1}{L} \sum_{i=1}^L K_{i,i} \right)$
\AlgLine{1}{1}
  \STATE $\vBPri = \vAExt = \left(\frac{1}{\vAPost} - \frac{1}{\vAPri}\right)^{-1}$
\AlgLine{1}{1}
  \STATE $\xBPri = \xAExt = \vAExt \, \left(\frac{\xAPost}{\vAPost} - \frac{\xAPri}{\vAPri}\right)$
      \label{alg_TSR_LineExtrA}
\AlgLine{0}{0}

\COMMENT{Soft feedback}
\AlgLine{1}{1}
  \STATE $\xBPosti = \E\{x_i|\xBPrii\} = \W(\xBPrii,\,\vBPri,\,s)$
      \label{algTSR_LineSF}
\AlgLine{1}{1}
  \STATE $\vBPost = \frac{1}{L} \sum_{i=1}^{L} \var\{x_i|\xBPrii\}$
\AlgLine{1}{1}
  \STATE $\vAPri = \vBExt = \left(\frac{1}{\vBPost} - \frac{1}{\vBPri}\right)^{-1}$
\AlgLine{1}{1}
  \STATE $\xAPri = \xBExt = \vBExt \, \left(\frac{\xBPost}{\vBPost} - \frac{\xBPri}{\vBPri}\right)$
  	\label{alg_TSR_LineExtrB}
\AlgLine{1}{1}
\ENDWHILE
\AlgLine{1}{1}	
\STATE $\hat{\ve{x}} = \quant_{\AO}(\xBPost)$
\end{algorithmic}
}
\end{algorithm}

\subsection{Exact LMMSE TSR}
\noindent
After the introduction of TSR/Q in the previous section, 
an improved version of the algorithm is introduced in this section. 
To this end, the true linear MMSE estimate for $\ve{x}$ is 
derived.
We assume the channel model (\ref{eq_channel_model}), i.e., 
$\ve{y} = \ve{A} \ve{x} + \ve{n}$, and want an estimate 
$\xAPost$ of $\ve{x}$ that minimizes the expectation of 
the squared error given prior knowledge on $\ve{x}$.
In general, the linear MMSE estimator is given by \cite{Kay:93, Sparrer:17}
\begin{align*}
\xAPost = \xAPri + \PhiDD \ve{A}^{\T} \left(\ve{A}\PhiDD\ve{A}^{\T} + \sigmaN^2 \ve{I}\right)^{-1} (\ve{y} - \ve{A}\xAPri) \ , 											
\end{align*}
where $\ve{d}$ is the error vector if $\xAPri = \ve{x} + \ve{d}$ 
is written as a noisy variant of $\ve{x}$, and 
$\PhiDD = \E\{\ve{d}\ve{d}^{\T}\}$ is its correlation matrix.  
If we assume that the elements of $\ve{d}$ are uncorrelated 
with variance $\vAPri$, the estimation can be 
simplified to
\begin{align*}
\xAPost = \xAPri + \ve{A}^{\T}\left(\ve{A}\ve{A}^{\T} + \frac{\sigmaN^2}{\vAPri} \ve{I}\right)^{-1} 
	    (\ve{y} - \ve{A}\xAPri) .
\end{align*}

The calculation of the soft values in the second step is 
equal to the one in TSR. 

This \ul{t}urbo-based algorithm, combining \ul{M}MSE 
estimation and \ul{s}oft feedback, is denoted by 
TMS/Q. The pseudocode of this algorithm 
is given in Alg.~\ref{Algo_TurboCS_withoutZ}, Variant B.

A comparison with the linear estimation in TSR/Q (cf.~(\ref{eq_ATSR_firstStepApprox})) 
shows that TSR/Q can be seen as a simplified version of TMS/Q, 
valid for $\ve{A}\ve{A}^{\T} \approx \c^2 \ve{I}$.
Note that TMS/Q holds for all types of measurement matrices,
the restriction that $\ve{A}$ is a (scaled) part of a unitary 
matrix, which was assumed in the original paper on TSR \cite{Ma:15a} 
and also for TSR/Q, is not required anymore.

%% file: TSR_bias_c3.tex
\section{Discussion of the Extrinsics Calculation}
\label{sec_3}
    
\begin{figure}
\centerline{\scalebox{0.62}{\input{Block_Diagram_joint.pstex_t}}}
\caption{\label{fig_BSB_TSR}
        Block diagram of TMS, interpreted from a turbo perspective (upper part), 
			      and interpreted from a signal theory perspective (lower part).
}
\end{figure}
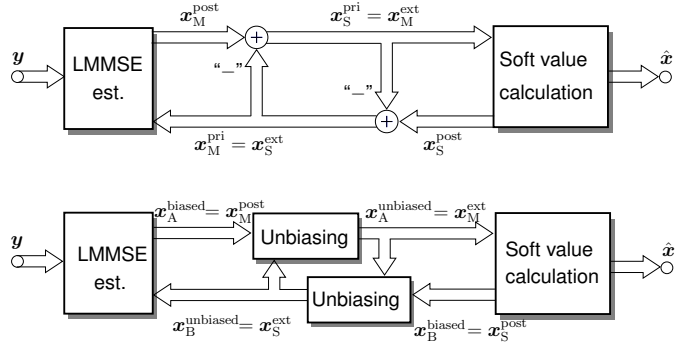

\noindent
The TMS algorithm is based on the turbo principle, which is well-known 
in channel coding \cite{Berrou:93}. The general idea is that two decoders 
(A and B in general, LMMSE estimator and soft value calculation here) 
exist, which alternately decode or estimate the signal and which exchange information 
about the current results. Each decoder takes into account 
the information from the other decoder (the so-called priors, i.e., $\xAPri$ \& $\xBPri$) 
and calculates a new estimate, the so-called posteriors ($\xAPost$ and $\xBPost$). In 
order not to hand back the information to the other decoder which has been received 
from it, this prior information has to be removed from the estimate (resulting in the 
so-called extrinsics, i.e., $\xAExt$ \& $\xBExt$) before passing it to the other decoder. The 
\emph{extrinsic values} are calculated by (cf.\ Line~\ref{alg_TSR_LineExtrA} \& \ref{alg_TSR_LineExtrB}, 
Alg.~\ref{Algo_TurboCS_withoutZ})
\begin{align}
	\xBPri = \xAExt = \vAExt \, \left(\frac{\xAPost}{\vAPost} - \frac{\xAPri}{\vAPri}\right) \ ,					\label{eq_unbiasingEstimates}
\end{align}
with corresponding average variances
\begin{align}
	\vBPri = \vAExt = \left(\frac{1}{\vAPost} - \frac{1}{\vAPri}\right)^{-1} \ .
\end{align}
This principle is shown in the upper part of Fig.~\ref{fig_BSB_TSR}.

In the following, both blocks of the algorithm are analyzed from a different point 
of view, leading to interesting insights.

\subsection{First Part}
\noindent
If discrete-valued signals are to be recovered, it is important that the diagonal 
elements of the end-to-end cascade for the estimation of $\ve{x}$ are equal 
to $1$, otherwise a \emph{bias} is present. In the MMSE case, the cascade is given by
\vspace*{-1mm}
\begin{align}
 \ve{K} = [K_{i,j}] = \ve{A}^{\T}\left(\ve{A}\ve{A}^{\T} + \frac{\sigmaN^2}{\vAPri} \ve{I}\right)^{-1}  \ve{A} \ ,
\end{align}
with diagonal elements smaller than one. 
In order to compensate for the bias, the estimates have to be scaled by the inverse 
of the diagonal elements of $\ve{K}$. Given the \emph{biased} estimate $\xAPri$, 
the $i^{\mathrm{th}}$ \emph{\ul{u}nbiased} element $\xUi$ is estimated by
\vspace*{-2mm}
\begin{align}
  \xUi &= \xAPrii + \frac{1}{K_{i,i}} \cdot (\xAPosti - \xAPrii) \notag \\
       &= \frac{1}{K_{i,i}} \cdot \xAPosti - \left(\frac{1}{K_{i,i}} - 1\right) \cdot \xAPrii \ .							\label{eq_xMMSEUnbiased}
\end{align}
With the prior error variance $\vAPri$, the \emph{biased} a-posteriori error 
variance of the $i^{\mathrm{th}}$ element is given by \cite{Kay:93} 
\begin{align}
\vAPosti &= \vAPri  - \vAPri \cdot \left[\ve{A}^{\T} \left(\ve{A}\ve{A}^{\T}+\frac{\sigmaN^2}{\vAPri}\right)^{-1} \ve{A}\right]_{(ii)} \notag \\
	 &= \vAPri \cdot (1-K_{i,i})	\ ,												\label{eq_varMMSEBiased}
\end{align}
and the \emph{unbiased} error variance $\sigmaUi$ calculates to \cite{Sparrer:17}
\begin{align}
\sigmaUi = \vAPri \cdot \left(\frac{1}{K_{i,i}}-1\right) \ .										\label{eq_varMMSEUnbiased}
\end{align}

Combining (\ref{eq_xMMSEUnbiased})--(\ref{eq_varMMSEUnbiased}), it follows
\begin{align}
\xUi &= \sigmaUi \cdot \left(\frac{\xAPosti}{\vAPosti} - \frac{\xAPrii}{\vAPri}\right)  \\ 
     &= \xAExti \\
\sigmaUi &= \vAExti \ ,
\end{align}
which is equal to the calculation of the extrinsic values (cf.\ Line~\ref{alg_TSR_LineExtrA}, 
Alg.~\ref{Algo_TurboCS_withoutZ}).
Hence, the extrinsic calculation corresponds to unbiasing 
of the estimate, and thus, when using unbiased MMSE estimates, (inherently) 
extrinsic information is considered.

\begin{figure}[ht]
\vspace*{-2mm}
\centerline{\scalebox{0.63}{\input{Vorward_Backward.pstex_t}}}
\caption{\label{fig_channelModels}
  {Channel models in the estimation process.}
}
\vspace*{-4mm}
\end{figure}
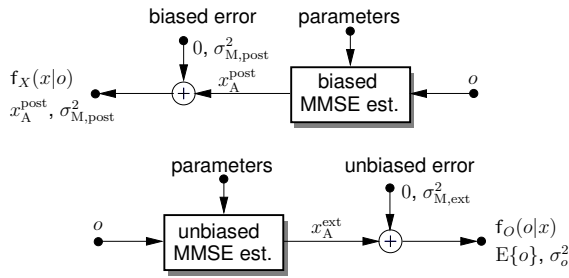

The unbiasing operation can also be interpreted from a third point 
of view. Given an observation $o$, the probability density function 
(pdf) of the biased a-posteriori estimate is given by
\begin{align}
\pdf_X(x|o) \sim \Gauss(\xAPosti,\vAPost) \ ,
\end{align}
which corresponds to the \emph{backward} channel model, cf.\ 
Fig.~\ref{fig_channelModels}, upper part. Note that in this 
model, as in any MMSE solution, the error is uncorrelated to the 
observation $o$. After the bias compensation, however, the error 
is uncorrelated to the unbiased estimate \cite{Fischer:02}, which 
leads to the \emph{forward} channel model which is shown in the 
lower part of Fig.~\ref{fig_channelModels}. 
In this case, the density of interest is $\pdf_O(o|x)$. 
With Bayes' theorem it holds 
\vspace*{-2mm}
\begin{align}
\pdf_O(o|x) = \pdf_O(o) \, \frac{\pdf_X(x|o)}{\pdf_X(x)} \ ,
\end{align}

\vspace*{-.5mm}
\noindent
where $\pdf_O(o)$ can be considered as a constant for a given $o$. If all 
variables are Gaussian distributed, it follows
\begin{align}
\pdf_O(o|x) \sim \, \frac{\Gauss(\xAPosti,\vAPost)}{\Gauss(\xAPrii,\vAPri)} \ ,
\end{align}

\vspace*{-.5mm}
\noindent
with the expected value and variance \cite{Guo:11}
\begin{align}
\textstyle
\xAExti = \E\{o|\xAPrii\} 
       = \vAExt \cdot \left(\frac{\xAPosti}{\vAPost} - \frac{\xAPrii}{\vAPri}\right) \\
\textstyle
\vAExt = \var\{o|\xAPrii\} 
       = \left(\frac{1}{\vAPost} - \frac{1}{\vAPri}\right)^{-1} \ ,
\end{align}

\vspace*{-.5mm}
\noindent
which are again the equations used for the extrinsic calculation in TMS. 
Thus, the \emph{biased} MMSE estimate is equal to the \emph{a-posteriori} value, 
which corresponds to the \emph{backward} channel model. The \emph{unbiased} MMSE 
estimate is equal to the \emph{extrinsic} value, and thus the \emph{forward} 
channel model.

\subsection{Second Part}
\noindent
In the second part of TMS, soft values are calculated. Since the estimator 
is non-linear, the calculation of the bias is not as obvious as for the 
first step. However, every non-linear memoryless device can be written as 
a linear scaling plus an additive estimation error which is uncorrelated 
to the linear part \cite{Banelli:13}. 
In general, any prior estimate $z$ of the transmitted value $x$ 
(with $z = \xBPrii$ in our case) can be written as noisy variant of $x$, 
i.e., $z = x + e$, where we assume the error to be Gaussian 
$e \sim \Gauss(0,\sigma_e^2)$, with $\sigma_e^2 = \vBPri$, and independent 
from $x$. The soft value of $z$ can be written as a function 
$\xBPosti= g(z) = \E\{x_i|z\}$, $z=\xBPrii$, which in turn 
is linearized as \cite{Banelli:13}
\vspace*{-1.5mm}
\begin{align}
z- g(z) = k_E e + n_E \ ,
\end{align}

\vspace*{-1mm}
\noindent
where the scaling factor
\vspace*{-1mm}
\begin{align}
\textstyle
k_E = \frac{\E_{ZE}\{(z-g(z))\, e\}}{\sigma_e^2} 
\end{align}

\vspace*{-1mm}
\noindent
is chosen such that the error $n_E$ has minimum variance. 
Note that, in contrast to the first step where we estimate the 
\emph{signal} $x$, the \emph{error} $e$ is estimated, with 
which in turn $x$ can be calculated. 
The estimated error after the soft feedback is given by 
$x+e-g(x+e)$, and hence 
\vspace*{-1mm}
\begin{align}
k_E &= \frac{\E_{XE}\{(x+e-g(x+e)) \, e\}}{\sigma_e^2}		\\
    &= \frac{\sigma_e^2 - \E\{g(x+e) \, e\}}{\sigma_e^2} \ .
\end{align}

\vspace*{-1mm}
\noindent
It can be shown via integration by parts that for any 
distribution of $x$, 
$\E\{g(x+e) \, e\} = \E\{g(\xBPrii) \, e\} = \vBPost = \E\{x_i^2|\xBPrii\} - (\E\{x_i|\xBPrii\})^2$ 
holds if the \emph{error} is Gaussian distributed.%
\footnote{
For a detailed derivation, please see the Appendix.
} 
Hence, taking $\sigma_e^2 = \vBPri$ into account, 
$k_E = \frac{\vBPri - \vBPost}{\vBPri}$, 
and with the resulting unbiasing factor $1/k_E$, 
the unbiased estimate calculates to 
$\xBExti = z - \frac{1}{k_E} \cdot (z-g(z)) = \xBPrii - \frac{\vBPri}{\vBPri - \vBPost} \cdot (\xBPrii-\xBPosti)$ 
and $\vBExt = \E\{(x-x_{\mathrm{S}}^{\mathrm{ext}})^2\}$, 
which, after straightforward modifications, results in the 
same unbiasing formulas as in the first step. Hence, as in 
the first step, the extrinsic calculation in TMS is equal 
to unbiasing, i.e., the change from the backward channel 
model to the forward one. The block diagram of TMS when it 
is interpreted from this unbiasing perspective is given in 
the lower part of Fig.~\ref{fig_BSB_TSR}

A comparison of both steps is given in Table~\ref{tab_comparisonOfSteps}. 
While the first step performs a vectorwise linear estimation of $\ve{x}$  
(implicitly assuming that $\ve{x}$ is Gaussian distributed), a non-linear 
symbolwise estimator for the error $\ve{e}$ is applied in the second step, 
thereby taking into account the sparsity and alphabet constraint.

\begin{table}
\caption{Comparison of the steps.
\vspace*{-.5mm}}
\label{tab_comparisonOfSteps}
\begin{tabular}{l||c|c}
\hline
			& First step				& Second step			\\ \hline \hline
Processing		& vectorwise (joint)			& symbolwise (individual)	\\ \hline
Assumption		& $x$ Gaussian distributed		& sparsity $s$, $x\in\AO$	\\ \hline
Estimation		& linear/affine				& non-linear			\\ \hline
Estimated variable	& signal $\ve{x}$			& error $\ve{e}$		\\ \hline

\end{tabular}
\vspace*{-2mm}
\end{table}

%% file: Block_Diagram_joint.pstex_t
\begin{picture}(0,0)%
\includegraphics{Block_Diagram_joint.pstex}%
\end{picture}%
\setlength{\unitlength}{4972sp}%
\begingroup\makeatletter\ifx\SetFigFont\undefined%
\gdef\SetFigFont#1#2#3#4#5{%
  \reset@font\fontsize{#1}{#2pt}%
  \fontfamily{#3}\fontseries{#4}\fontshape{#5}%
  \selectfont}%
\fi\endgroup%
\begin{picture}(5276,2734)(-21,-5360)
\put(1967,-4557){\makebox(0,0)[lb]{\smash{{\SetFigFont{12}{14.4}{\sfdefault}{\mddefault}{\updefault}{\color[rgb]{0,0,0}Unbiasing}%
}}}}
\put(2376,-5052){\makebox(0,0)[lb]{\smash{{\SetFigFont{12}{14.4}{\sfdefault}{\mddefault}{\updefault}{\color[rgb]{0,0,0}Unbiasing}%
}}}}
\put(631,-3391){\makebox(0,0)[lb]{\smash{{\SetFigFont{12}{14.4}{\sfdefault}{\mddefault}{\updefault}{\color[rgb]{0,0,0}est.}%
}}}}
\put(2881,-2761){\makebox(0,0)[b]{\smash{{\SetFigFont{12}{14.4}{\sfdefault}{\mddefault}{\updefault}{\color[rgb]{0,0,0}$\xBPri=\xAExt$}%
}}}}
\put(1486,-2761){\makebox(0,0)[b]{\smash{{\SetFigFont{12}{14.4}{\sfdefault}{\mddefault}{\updefault}{\color[rgb]{0,0,0}$\xAPost$}%
}}}}
\put(1801,-3796){\makebox(0,0)[b]{\smash{{\SetFigFont{12}{14.4}{\sfdefault}{\mddefault}{\updefault}{\color[rgb]{0,0,0}$\xAPri=\xBExt$}%
}}}}
\put(3421,-3796){\makebox(0,0)[b]{\smash{{\SetFigFont{12}{14.4}{\sfdefault}{\mddefault}{\updefault}{\color[rgb]{0,0,0}$\xBPost$}%
}}}}
\put(2746,-3391){\makebox(0,0)[b]{\smash{{\SetFigFont{12}{14.4}{\sfdefault}{\mddefault}{\updefault}{\color[rgb]{0,0,0}``$-$''}%
}}}}
\put(1711,-3211){\makebox(0,0)[b]{\smash{{\SetFigFont{12}{14.4}{\sfdefault}{\mddefault}{\updefault}{\color[rgb]{0,0,0}``$-$''}%
}}}}
\put(473,-3166){\makebox(0,0)[lb]{\smash{{\SetFigFont{12}{14.4}{\sfdefault}{\mddefault}{\updefault}{\color[rgb]{0,0,0}LMMSE}%
}}}}
\put(646,-4917){\makebox(0,0)[lb]{\smash{{\SetFigFont{12}{14.4}{\sfdefault}{\mddefault}{\updefault}{\color[rgb]{0,0,0}est.}%
}}}}
\put(511,-4692){\makebox(0,0)[lb]{\smash{{\SetFigFont{12}{14.4}{\sfdefault}{\mddefault}{\updefault}{\color[rgb]{0,0,0}LMMSE}%
}}}}
\put(3871,-3391){\makebox(0,0)[lb]{\smash{{\SetFigFont{12}{14.4}{\sfdefault}{\mddefault}{\updefault}{\color[rgb]{0,0,0}calculation}%
}}}}
\put(3871,-3121){\makebox(0,0)[lb]{\smash{{\SetFigFont{12}{14.4}{\sfdefault}{\mddefault}{\updefault}{\color[rgb]{0,0,0}Soft value}%
}}}}
\put(3916,-4647){\makebox(0,0)[lb]{\smash{{\SetFigFont{12}{14.4}{\sfdefault}{\mddefault}{\updefault}{\color[rgb]{0,0,0}Soft value}%
}}}}
\put(3916,-4870){\makebox(0,0)[lb]{\smash{{\SetFigFont{12}{14.4}{\sfdefault}{\mddefault}{\updefault}{\color[rgb]{0,0,0}calculation}%
}}}}
\put(1566,-4339){\makebox(0,0)[b]{\smash{{\SetFigFont{12}{14.4}{\sfdefault}{\mddefault}{\updefault}{\color[rgb]{0,0,0}$\ve{x}_{\mathrm{A}}^{\mathrm{biased}} \hspace*{-1.5mm} = \xAPost$}%
}}}}
\put(3286,-4341){\makebox(0,0)[b]{\smash{{\SetFigFont{12}{14.4}{\sfdefault}{\mddefault}{\updefault}{\color[rgb]{0,0,0}$\ve{x}_{\mathrm{A}}^{\mathrm{unbiased}} \hspace*{-1mm} = \xAExt$}%
}}}}
\put(3642,-5291){\makebox(0,0)[b]{\smash{{\SetFigFont{12}{14.4}{\sfdefault}{\mddefault}{\updefault}{\color[rgb]{0,0,0}$\ve{x}_{\mathrm{B}}^{\mathrm{biased}} \hspace*{-1mm} = \xBPost$}%
}}}}
\put(1744,-5250){\makebox(0,0)[b]{\smash{{\SetFigFont{12}{14.4}{\sfdefault}{\mddefault}{\updefault}{\color[rgb]{0,0,0}$\ve{x}_{\mathrm{B}}^{\mathrm{unbiased}} \hspace*{-1.5mm} = \xBExt$}%
}}}}
\put( 43,-3082){\makebox(0,0)[b]{\smash{{\SetFigFont{12}{14.4}{\sfdefault}{\mddefault}{\updefault}{\color[rgb]{0,0,0}$\ve{y}$}%
}}}}
\put( 31,-4563){\makebox(0,0)[b]{\smash{{\SetFigFont{12}{14.4}{\sfdefault}{\mddefault}{\updefault}{\color[rgb]{0,0,0}$\ve{y}$}%
}}}}
\put(5215,-4647){\makebox(0,0)[b]{\smash{{\SetFigFont{12}{14.4}{\sfdefault}{\mddefault}{\updefault}{\color[rgb]{0,0,0}$\hat{\ve{x}}$}%
}}}}
\put(5200,-3112){\makebox(0,0)[b]{\smash{{\SetFigFont{12}{14.4}{\sfdefault}{\mddefault}{\updefault}{\color[rgb]{0,0,0}$\hat{\ve{x}}$}%
}}}}
\end{picture}%

%% file: Vorward_Backward.pstex_t
\begin{picture}(0,0)%
\includegraphics{Vorward_Backward.pstex}%
\end{picture}%
\setlength{\unitlength}{4972sp}%
\begingroup\makeatletter\ifx\SetFigFont\undefined%
\gdef\SetFigFont#1#2#3#4#5{%
  \reset@font\fontsize{#1}{#2pt}%
  \fontfamily{#3}\fontseries{#4}\fontshape{#5}%
  \selectfont}%
\fi\endgroup%
\begin{picture}(3831,2132)(-610,-17)
\put(687,104){\makebox(0,0)[lb]{\smash{{\SetFigFont{12}{14.4}{\sfdefault}{\mddefault}{\updefault}{\color[rgb]{0,0,0}MMSE est.}%
}}}}
\put(723,280){\makebox(0,0)[lb]{\smash{{\SetFigFont{12}{14.4}{\sfdefault}{\mddefault}{\updefault}{\color[rgb]{0,0,0}unbiased}%
}}}}
\put(2472,592){\makebox(0,0)[lb]{\smash{{\SetFigFont{12}{14.4}{\sfdefault}{\mddefault}{\updefault}{\color[rgb]{0,0,0}$0$, $\vAExt$}%
}}}}
\put(1678,1280){\makebox(0,0)[lb]{\smash{{\SetFigFont{12}{14.4}{\sfdefault}{\mddefault}{\updefault}{\color[rgb]{0,0,0}MMSE est.}%
}}}}
\put(1819,1460){\makebox(0,0)[lb]{\smash{{\SetFigFont{12}{14.4}{\sfdefault}{\mddefault}{\updefault}{\color[rgb]{0,0,0}biased}%
}}}}
\put(670,804){\makebox(0,0)[lb]{\smash{{\SetFigFont{12}{14.4}{\sfdefault}{\mddefault}{\updefault}{\color[rgb]{0,0,0}parameters}%
}}}}
\put(1666,1980){\makebox(0,0)[lb]{\smash{{\SetFigFont{12}{14.4}{\sfdefault}{\mddefault}{\updefault}{\color[rgb]{0,0,0}parameters}%
}}}}
\put(2026,802){\makebox(0,0)[lb]{\smash{{\SetFigFont{12}{14.4}{\sfdefault}{\mddefault}{\updefault}{\color[rgb]{0,0,0}unbiased error}%
}}}}
\put(1036,1476){\makebox(0,0)[lb]{\smash{{\SetFigFont{12}{14.4}{\sfdefault}{\mddefault}{\updefault}{\color[rgb]{0,0,0}$x_{\mathrm{A}}^{\mathrm{post}}$}%
}}}}
\put( 58,341){\makebox(0,0)[lb]{\smash{{\SetFigFont{12}{14.4}{\sfdefault}{\mddefault}{\updefault}{\color[rgb]{0,0,0}$o$}%
}}}}
\put(-595,1241){\makebox(0,0)[lb]{\smash{{\SetFigFont{12}{14.4}{\sfdefault}{\mddefault}{\updefault}{\color[rgb]{0,0,0}$x_{\mathrm{A}}^{\mathrm{post}}$, $\vAPost$}%
}}}}
\put(2993,1493){\makebox(0,0)[lb]{\smash{{\SetFigFont{12}{14.4}{\sfdefault}{\mddefault}{\updefault}{\color[rgb]{0,0,0}$o$}%
}}}}
\put(1764,316){\makebox(0,0)[lb]{\smash{{\SetFigFont{12}{14.4}{\sfdefault}{\mddefault}{\updefault}{\color[rgb]{0,0,0}$x_{\mathrm{A}}^{\mathrm{ext}}$}%
}}}}
\put(3206, 95){\makebox(0,0)[lb]{\smash{{\SetFigFont{12}{14.4}{\sfdefault}{\mddefault}{\updefault}{\color[rgb]{0,0,0}$\E\{o\}$, $\sigma_o^2$}%
}}}}
\put(3203,321){\makebox(0,0)[lb]{\smash{{\SetFigFont{12}{14.4}{\sfdefault}{\mddefault}{\updefault}{\color[rgb]{0,0,0}$\pdf_O(o|x)$}%
}}}}
\put(-595,1496){\makebox(0,0)[lb]{\smash{{\SetFigFont{12}{14.4}{\sfdefault}{\mddefault}{\updefault}{\color[rgb]{0,0,0}$\pdf_X(x|o)$}%
}}}}
\put(837,1736){\makebox(0,0)[lb]{\smash{{\SetFigFont{12}{14.4}{\sfdefault}{\mddefault}{\updefault}{\color[rgb]{0,0,0}$0$, $\vAPost$}%
}}}}
\put(496,1976){\makebox(0,0)[lb]{\smash{{\SetFigFont{12}{14.4}{\sfdefault}{\mddefault}{\updefault}{\color[rgb]{0,0,0}biased error}%
}}}}
\end{picture}%

%% file: TSR_bias_c4.tex
\section{Simulation Results}
\label{sec_4}

\begin{figure}
\psfrag{SER}[Bc][Bc]{$\SER \ \rightarrow$}
\psfrag{SNR}[cc][Bc]{$10 \log_{10} (1/\sigmaN^2)  \; [\dB] \ \rightarrow$}
\psfrag{ABBBBBB}{\hspace*{-4mm}\scriptsize IHT/Q}
\psfrag{BABBBBB}{\hspace*{-4mm}\scriptsize ISFT/Q}
\psfrag{BBABBBB}{\hspace*{-4mm}\scriptsize BAMP/Q}
\psfrag{BBBABBB}{\hspace*{-4mm}\scriptsize TSR/Q}
\psfrag{BBBBABB}{\hspace*{-4mm}\scriptsize TMS/Q}
\psfrag{BBBBBAB}{\hspace*{-4mm}\scriptsize IMS/Q}
 
 \includegraphics[width=.98\columnwidth]{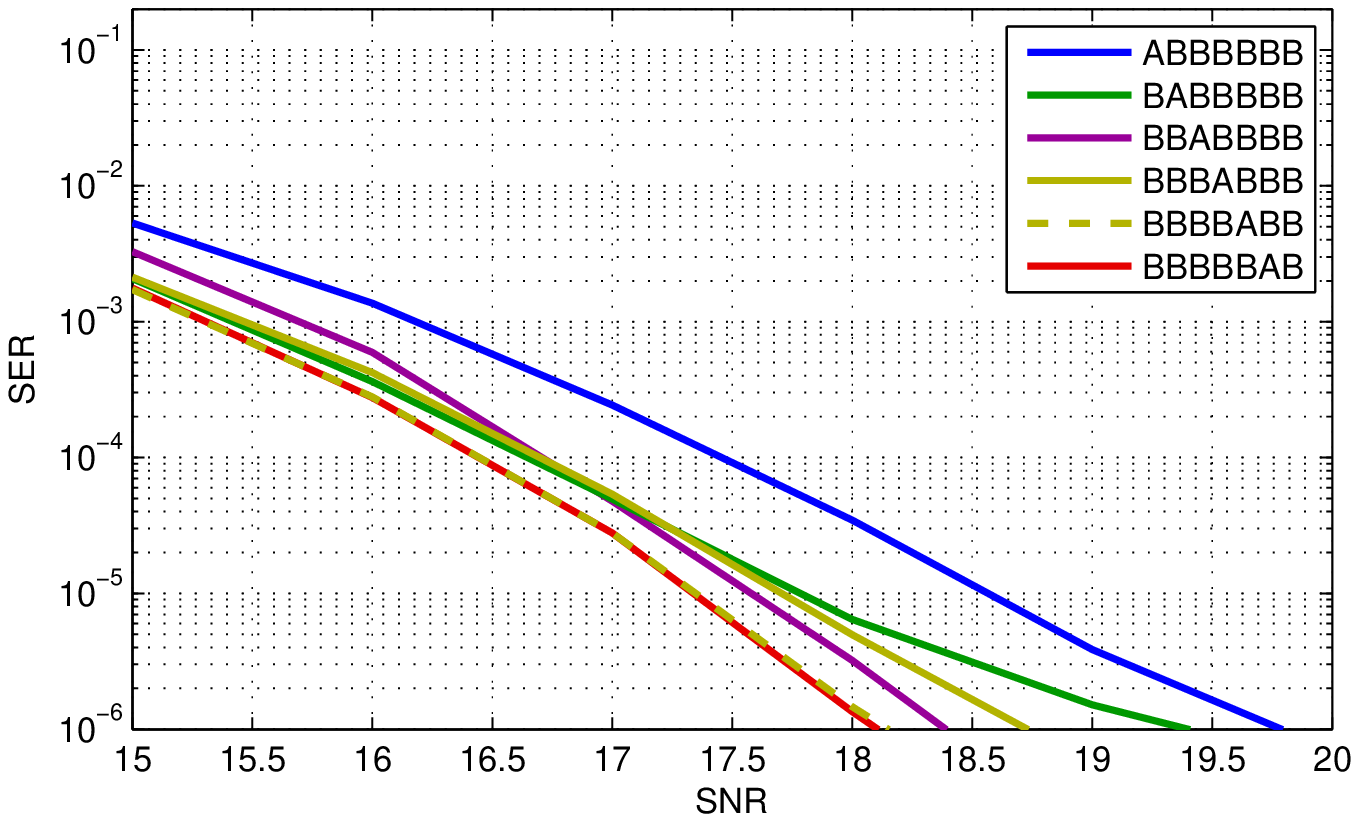}

\vspace*{.3em}
 \includegraphics[width=.98\columnwidth]{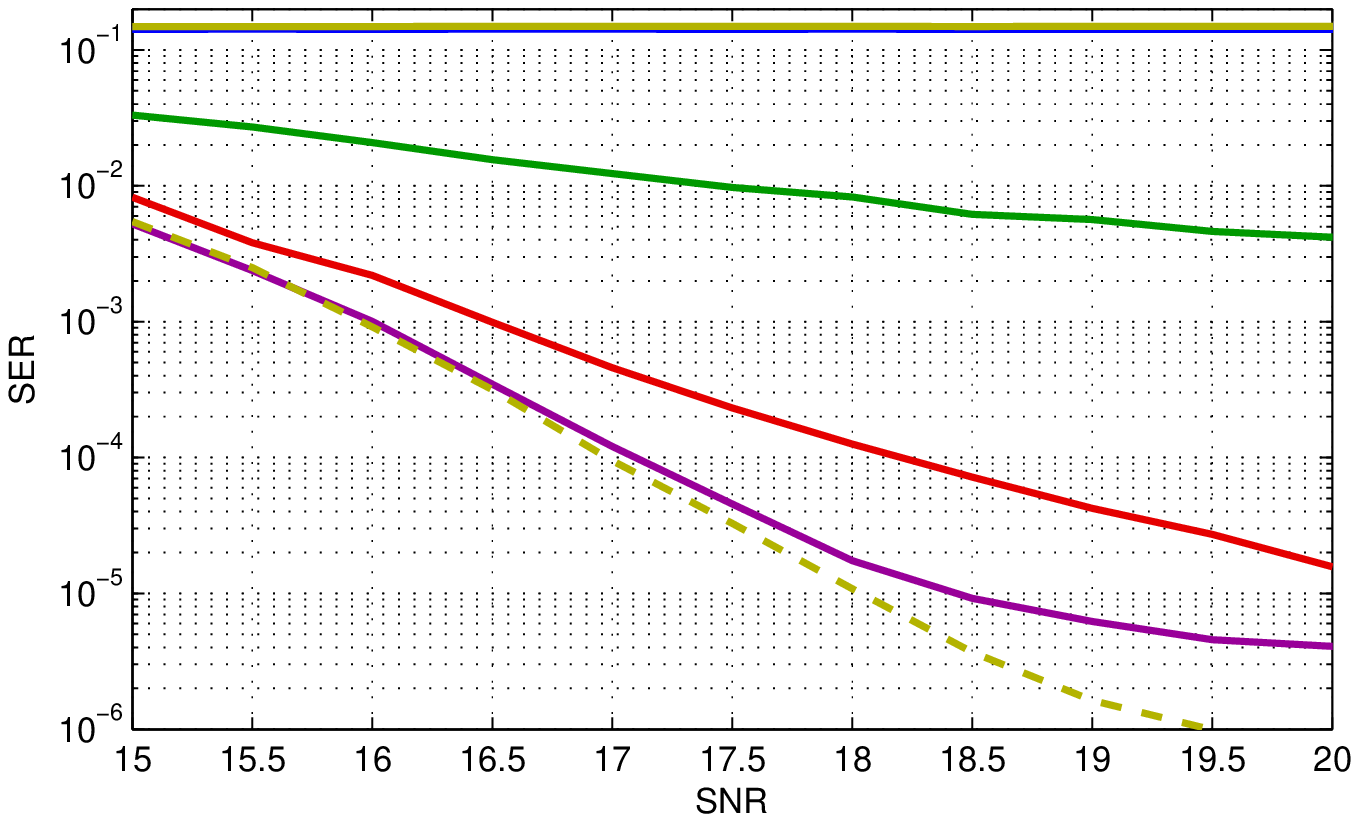}
 \caption{\label{fig_SER}
 	$\SER$ over the noise level
 	$1/\sigma_n^2$ in $\dB$. $L=258$, $K=129$, $s=20$, $\A=\{-1, +1\}$.
 	SVD-based matrix (top), Gaussian matrix (bottom). $\|\ve{A}_{(:,i)}\|^2=1\ \forall \ i$.
 \vspace*{-2mm}}
 \end{figure}

\noindent
In this section, numerical results of the new algorithm are shown 
and compared to the ones of established algorithms. 
Two different channel matrices are used. First, $\ve{A}$ is constructed as 
random part of a random orthogonal matrix. Second, $\ve{A}$ is a random 
Gaussian matrix. In both cases, the columns are normalized to unit length, 
and $L=258$, $K=129$, $s=20$, $\AO=\{-1,0,+1\}$. To ensure convergence, 
all algorithms perform $50$ iterations. The measure of interest when 
dealing with discrete values is the {s}ymbol \ul{e}rror \ul{r}ate 
$\SER = \frac{1}{L} \sum_{i=1}^L \Pr\{\hat{x}_i \neq x_i\}$. 
The results are shown in Fig.~\ref{fig_SER}. 
In the SVD-based case (upper figure), the new algorithm (yellow, dashed) 
performs as good as IMS/Q (red) \cite{Sparrer:17}. TSR/Q (yellow, solid) 
is outperformed by $0.5\,\dB$, and the well-known BAMP/Q by $0.2\,\dB$. 
For comparison, the results of the standard algorithms for discrete CS 
(IHT/Q, blue, and ISFT/Q, green) are also shown.

For Gaussian sensing matrices, the new algorithm TMS/Q shows the best 
performance. Note that TSR/Q fails since the assumption that $\ve{A}$ 
is a part of an orthogonal matrix is not fulfilled. In general, the 
Gaussian matrices are less structured than the SVD-based,
and only a smaller sparsity is tolerated to obtain the same 
performance. Note that the influence of the matrix is much larger for 
algorithms in which the matched filter is applied (IHT/Q, ISFT/Q) than 
for MMSE-estimation-based algorithms (TMS/Q, IMS/Q). However, also the 
aforementioned algorithms are able to find the correct result if the 
sparsity is small enough.

%% file: TSR_bias_c5.tex
\section{Conclusion}
\label{sec_5}

\noindent
In this paper, we have shown that the calculation of 
extrinsic information is equal to unbiasing, justifying 
the calculation performed in the second step of TSR and TMS. 
In addition, the TSR algorithm has been improved by the 
application of the LMMSE estimation, which leads to  better 
numerical results than the established algorithms.

%% file: TSR_bias_a1.tex
\begin{appendix}

\noindent
In the following, we proof that $\E\nolimits_{X,E}\{g(x+e) \cdot e\} = \vBPost$. 
Since we assume $x$ and $e$ in the channel model $z=x+e$ to be independent, we have
\begin{align}
&\E\nolimits_{X,E}\{g(x+e) \cdot e\} = \int \int g(x+e) \, e \, \pdf_X(x)  \, \pdf_E(e) \de \dx				\notag		\\ 
  &\hspace*{1cm}= \int \pdf_X(x) \, \left(\int (z-x) \, \pdf_E(z-x) \, g(z) \dz \right) \dx				\notag 		\\
  &\hspace*{1cm}\defeq \int \pdf_X(x) \, I_1(x) \dx  									\label{eq_I1}	\\
&\vBPost = \E\nolimits_Z\{\vBPosti(z)\} = \E\nolimits_{X,E}\{\vBPosti(x+e)\} 						\notag 		\\
  &\hspace*{1cm}= \int \int \vBPosti(x+e) \, \pdf_X(x) \pdf_E(e) \de \dx 						\notag 		\\
  &\hspace*{1cm}= \int \pdf_X(x) \left( \int \pdf_E(z-x) \, \vBPosti(z) \dz \right) \dx 				\notag 		\\
  &\hspace*{1cm}\defeq \int \pdf_X(x) \, I_2(x) \dx	\ .								\label{eq_I2}
\end{align}
In order to proof that $\E\{e \cdot g(x+e)\} = \vBPost$ holds, 
we show that $I_1(x) = I_2(x)$, 
which is a stricter condition than to proof ``only'' the equality 
of the entire integrals.

In the following, we replace the variable $x$ of the integral 
by $q$ to avoid confusion. With $g(z) = \E\{x|z\} = \int x \, \pdf_X(x|z) \dx$,
$\vBPosti = \E\{x^2|z\} - (\E\{x|z\})^2 = \int x^2 \, \pdf_X(x|z) \dx - (\int x \, \pdf_X(x|z) \dx)^2$, 
$f_Z(z|x) = f_E(z-x)$, and Bayes' theorem follows%
\footnote{
If no limits are given for integrals, the lower and upper limits 
are $-\infty$ and $\infty$, respectively.
}
\begin{align*}
I_1(q) &= \int (z-q) \ \pdf_{E}(z-q) \ \frac{\int x \pdf_E(z-x) \pdf_X(x) \dx}{\pdf_Z(z)} \dz 			\\
I_2(q) &= \int \pdf_E(z-q) \ \frac{\int x^2 \pdf_E(z-x) \pdf_X(x) \dx}{\pdf_Z(z)} \dz 				\\
       & - \int \pdf_E(z-q) \ \bigg(\frac{\int x \pdf_E(z-x) \pdf_X(x) \dx}{f_Z(z)}\bigg)^2 \dz 		\ .
\end{align*}
In the following, we assume that the error is Gaussian distributed 
with variance $\sigma_e^2$, i.e., $\pdf_{E}(e) = c \e^{-e^2/(2\sigma_e^2)}$, 
$c=\frac{1}{\sqrt{2\pi\sigma_e^2}}$. 
We reformulate $I_1(q)$ via integration in parts. It holds
\begin{align}
\int f'(z) g(z) \dz &= [f(z)g(z)]_{-\infty}^{\infty} - \int g'(z) f(z) \dz 					\notag			\\
		    &= - \int g'(z) f(z) \dz								\ ,	\label{eq_PartIntegral}
\end{align}
with
\begin{align}
f'(z) &= (z-q) \ \pdf_{E}(z-q) = (z-q) c \e^{-(z-q)^2/(2\sigma_e^2)} 						\label{eq_fStrich} 	\\
f(z)  &= - \sigma_e^2 c \e^{-(z-q)^2/(2\sigma_e^2)} 								\label{eq_f} 		\\
g(z)  &= \frac{\int x \pdf_E(z-x) \pdf_X(x) \dx}{\pdf_Z(z)} 
        \defeq \frac{u(z)}{v(z)} 										\label{eq_g} 		\\
g'(z) &= \frac{u'(z) v(z) - u(z) v'(z)}{(v(z))^2} 						\ .		\label{eq_gStrich}
\end{align}
Note that $[f(z)g(z)]_{-\infty}^{\infty} = 0$ since $\lim_{|z|\to\infty} f(z) = 0$. 
For Gaussian error, the factors in (\ref{eq_gStrich}) are given by
\begin{align*}
u(z)  &= \int x \pdf_E(z-x) \pdf_X(x) \dx 										\\
      &= \int x \, c \, \e^{-(z-x)^2/(2\sigma_e^2)} \pdf_X(x) \dx 							\\         
u'(z) &= \frac{\d}{\dz} \int x \, c \, \e^{-(z-x)^2/(2\sigma_e^2)} \pdf_X(x) \dx 					\\
      &= - \int x \ c \ \frac{1}{\sigma_e^2} (z-x) \e^{-(z-x)^2/(2\sigma_e^2)} \pdf_X(x) \dx 				\\
      &= - \frac{1}{\sigma_e^2} \, \int x \, (z-x) \, \pdf_E(z-x) \pdf_X(x) \dx 					\\ 
v(z)  &= \pdf_Z(z) 													\\
      &= \int \pdf_E(z-x) \pdf_X(x) \dx 										\\ 
      &= \int c \, \e^{-(z-x)^2/(2\sigma_e^2)} \pdf_X(x) \dx 								\\
v'(z) &= \frac{\d}{\dz} \int c \e^{-(z-x)^2/(2\sigma_e^2)} \pdf_X(x) \dx 						\\
      &= - \int c \, \frac{1}{\sigma_e^2} \, (z-x) \, \e^{-(z-x)^2/(2\sigma_e^2)} \pdf_X(x) \dx 			\\
      &= - \frac{1}{\sigma_e^2} \, \int (z-x) \, \pdf_E(z-x) \pdf_X(x) \dx 						\\
      &= - \frac{1}{\sigma_e^2} \, z \int \pdf_E(z-x) \pdf_X(x) \dx 						\notag 	\\
      &\hphantom{=\ } + \frac{1}{\sigma_e^2} \, \int x \, \pdf_E(z-x) \pdf_X(x) \dx 				\ .      
\end{align*}
Hence $g'(z)$ calculates to
\begin{align}
g'(z) &= \frac{u'(z) v(z) - u(z) v'(z)}{(v(z))^2} 								\notag	\\
      &= - \frac{1}{(v(z))^2}  \cdot \frac{1}{\sigma_e^2} \, \cdot						\notag	\\
	&\hphantom{=\ }  \bigg( z \cdot \int x \, \pdf_E(z-x) \pdf_X(x) \dx \cdot \int \pdf_E(z-x) \pdf_X(x) \dx\notag	\\
	&\hphantom{=\ } - \int x^2 \, \pdf_E(z-x) \pdf_X(x) \dx \cdot \int \pdf_E(z-x) \pdf_X(x) \dx 		\notag	\\
        &\hphantom{=\ } - z \cdot \int \pdf_E(z-x) \pdf_X(x) \dx \cdot \int x \, \pdf_E(z-x) \pdf_X(x) \dx 	\notag	\\
	&\hphantom{=\ } + \int x \, \pdf_E(z-x) \pdf_X(x) \dx \cdot \int x \, \pdf_E(z-x) \pdf_X(x) \dx \bigg) 	\notag	\\
      &= \frac{1}{(v(z))^2}  \cdot \frac{1}{\sigma_e^2} \, \cdot				\label{eq_gStrich2}	\\
	&\hphantom{=\ } \bigg( \int x^2 \, \pdf_E(z-x) \pdf_X(x) \dx \cdot \int \pdf_E(z-x) \pdf_X(x) \dx 	\notag	\\
	&\hphantom{=\ } - \int x \, \pdf_E(z-x) \pdf_X(x) \dx \cdot \int x \, \pdf_E(z-x) \pdf_X(x) \dx \bigg) 	\notag
\end{align}
Plugging (\ref{eq_fStrich})--(\ref{eq_g}) and (\ref{eq_gStrich2}) 
into (\ref{eq_PartIntegral}) yields
\begin{align*}
I_1(q) &= - \int f(z) \, g'(z) \dz
\end{align*}
\begin{align*}
    &= \int \sigma_e^2 \, c \, \e^{-(z-q)^2/(2\sigma_e^2)} \cdot \frac{1}{(\pdf_Z(z))^2 \cdot \sigma_e^2}  \cdot 		\\
	&\hphantom{=\ } \bigg( \int x^2 \, \pdf_E(z-x) \pdf_X(x) \dx \cdot \int \pdf_E(z-x) \pdf_X(x) \dx			\\
	&\hphantom{= \bigg(} - \left(\int x \, \pdf_E(z-x) \pdf_X(x) \dx\right)^2 \bigg) \dz 					\\
    &= \int \pdf_E(z-q) \cdot  \bigg(\frac{\int x^2 \, \pdf_E(z-x) \pdf_X(x) \dx} {\pdf_Z(z)} \bigg) \dz		\notag	\\
	&\hphantom{=\ } - \int \pdf_E(z-q) \cdot \bigg(\frac{\int x \, \pdf_E(z-x) \pdf_X(x) \dx}{\pdf_Z(z)} \bigg)^2 \dz	\\
    &= I_2(q) \ .
\end{align*}
Thus, both integrals are equal, and hence 
$\E\nolimits_{X,E}\{g(x+e) \cdot e\} = \vBPost$. 

Note that the unbiasing is independent of the distribution $\pdf_X(x)$, 
i.e., it holds for all possible alphabets, hence for discrete, and 
also for continuous (real-valued) distributions of the elements of the 
sparse vector. The only condition is that the \emph{error} is Gaussian 
distributed.

\end{appendix}